# Artificial Intelligence-Assisted Systems Biology Reconstruction of Diabetic Bone Marrow Dysfunction

**Running title:** Systems Biology of Diabetic Bone Marrow

**Paolo Madeddu, MD**

**Correspondence:**
Professor Paolo Madeddu, MD
Bristol Medical School
University of Bristol
Bristol BS8 1TH
United Kingdom
Email: mdprm@bristol.ac.uk

**Word count**

Main text: **approximately 7,750** words
Abstract: **237** words

**Tables:** 10

**Figures: 6**

**Supplementary Material**

Supplementary File S1 — Final reconstruction corpus publication list
Supplementary File S2 — Corpus certification and signed freeze certificate
Supplementary File S3 — Full verbatim prompts used for LLM interrogation
Supplementary File S4 — Discovery concordance analysis
Supplementary File S5 — Validation concordance analysis
Supplementary File S6 — Final synthesis concordance analysis
Supplementary File S7 — AI-assisted meta-analysis
Supplementary File S8 — Consensus Biological Atlas (Cross Reasoning Atlas)
Supplementary File S9 — Node dictionary
Supplementary File S10 — Directed edge table
Supplementary File S11 — Original Gephi network screenshot
Supplementary File S12 — Endothelial deletion screenshot
Supplementary File S13 — External plausibility audit
Supplementary File S14 — Physiological reference audit trail
Supplementary File S15 — Augmented-edge documentation
Supplementary File S16 — Interrogation-fidelity audit

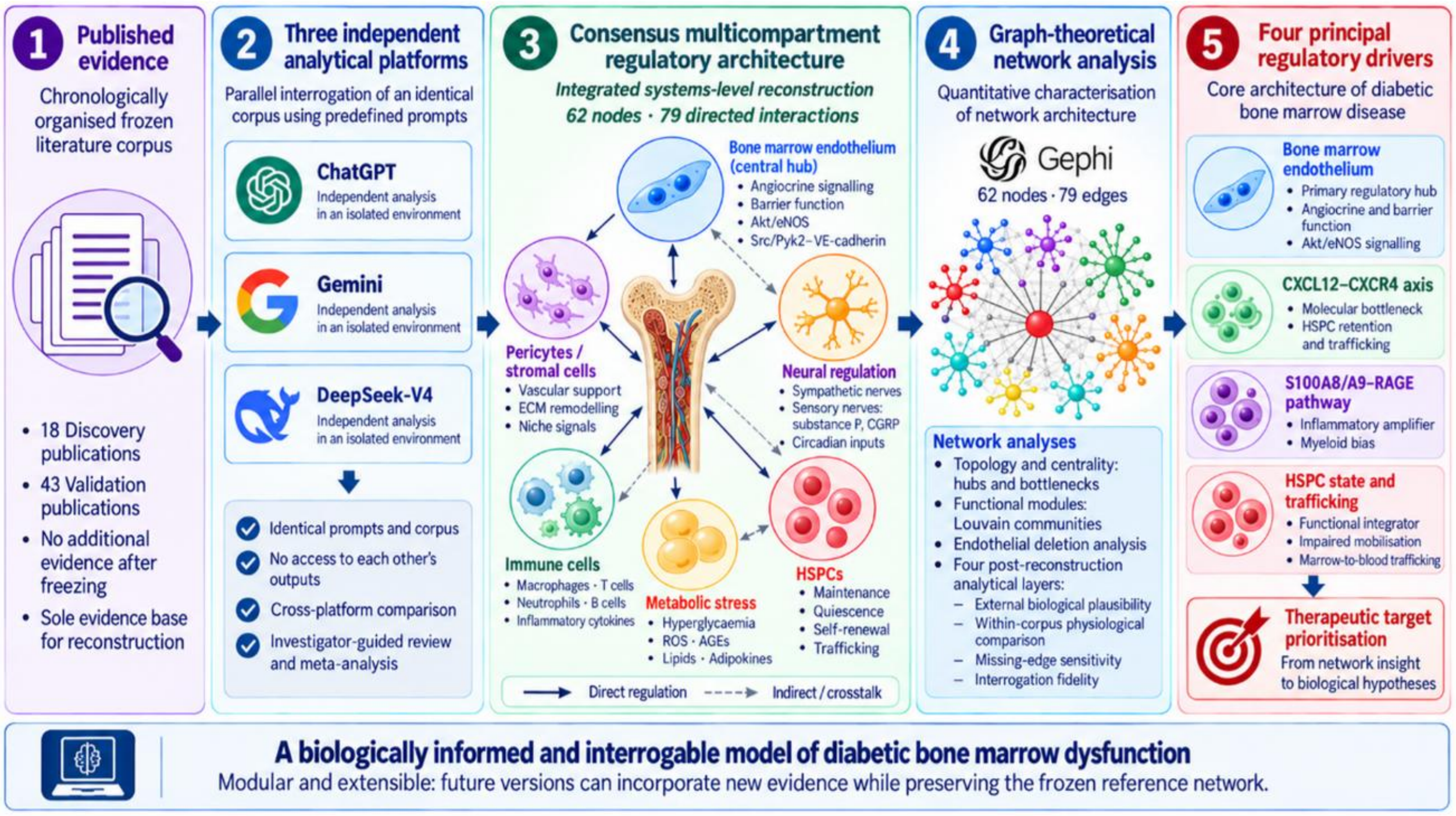

**Graphical Abstract.** AI-assisted systems biology reconstruction of diabetic bone marrow dysfunction. Three large language models analysed a frozen corpus of 18 Discovery and 43 Validation publications. Investigator-guided synthesis produced a consensus network of 62 nodes and 79 directed interactions. Topology, endothelial deletion and four post-reconstruction assessments examined its organisation, biological plausibility and interrogation fidelity. The model connects niche regulation, inflammatory signalling and haematopoietic stem and progenitor cell state and trafficking, identifying relationships for experimental testing. New evidence can inform separately versioned reconstructions.

# ABSTRACT

**Aims/hypothesis.** Diabetes disrupts bone marrow niches, but the high-level understanding of this disruption remains poorly defined. This study aimed to reconstruct a regulatory network suitable for systems biology analysis of diabetic bone marrow.

**Methods.** Three artificial intelligence large language models independently analysed 18 Discovery and 43 Validation publications. Investigator-guided integration produced a frozen network of 62 biological entities and 79 directed interactions. Open-source software Gephi created a visual representation before and after main node removal. The plausibility of the reconstructed interactions was tested against independent curated knowledge resources, including the Kyoto Encyclopedia of Genes and Genomes, connectomeDB2025, Reactome, and Gene Ontology. Further analysis examined physiological correspondence, missing-edge sensitivity, and graph-constrained interrogation.

**Results.** Functional modules encompassed endothelial–perivascular support, stem cell retention and trafficking, neural–circadian regulation, inflammatory myelopoiesis, metabolic stress, and epigenetic regulation. They intersected through shared effects on endothelial function and stem-cell behaviour. The endothelium was the principal connecting node. Computational removal of the endothelial node increased network fragmentation and diameter, while average directed path length remained essentially unchanged. External resources supported 60/79 interactions (75.9%); 19 remained unresolved, with none classified as contradictory. Complementary assessments supported biological plausibility and faithful interrogation of the reconstructed network.

**Conclusions/interpretation.** This investigator-led dialogue with artificial intelligence provides a foundation for a systems-level understanding of diabetic bone marrow dysfunction. The approach may facilitate the generation of testable hypotheses and the identification of mechanistic targets for coordinated restoration of disrupted niche interactions.

# INTRODUCTION

Over the past two decades, the bone marrow (BM) has been recognised not only as the primary site of haematopoiesis but also as a regulator of tissue homeostasis and regeneration. These functions depend on anatomically and functionally distinct niches connected through cellular and molecular interactions.[1–3] Stromal support and stem-cell maintenance,[4–7] vascular and angiocrine signalling,[8–15] neural and circadian control of cell trafficking,[16–19] and adipocyte and immune regulation[20–22] collectively shape local bone marrow homeostasis. Studies of circulating progenitor cells also link this organisation to vascular repair.[23,24] Niche heterogeneity and communication across compartments therefore provide a biological basis for a systems-level understanding of BM responses to local injury and systemic disease.[25–29]

Diabetes mellitus illustrates how systemic disease can disrupt this organisation. Early studies identified impaired function of circulating endothelial progenitor cells,[30,31] while subsequent work established diabetic stem cell mobilopathy: defective mobilisation of haematopoietic stem and progenitor cells (HSPCs) from the marrow.[32,49] The *emerging picture* extended beyond intrinsic progenitor cell defects to include microangiopathy and endothelial barrier disruption,[33–35] autonomic and sensory neuropathy,[37,38] altered trafficking signals,[39,40] pericyte dysfunction,[42] inflammatory activation[43,44] and adipogenic stromal remodelling.[45] Evidence from experimental and clinical studies further linked diabetes to impaired progenitor cell availability.[48–50] Other studies identified altered endothelial control of haematopoiesis and increased inflammatory myeloid output.[52,58] These findings support an integrated view of diabetic BM dysfunction, while its contribution to individual peripheral complications remains a distinct question.[41,61,62]

Despite these advances, the systems-level organisation of diabetic BM disease remains incompletely defined. Its cellular and molecular mechanisms are increasingly recognised as interdependent, but much of the evidence derives from studies of individual components conducted under different experimental conditions and using heterogeneous phenotypic definitions. Consequently, a substantial body of mechanistic knowledge has accumulated without an integrated representation of the direction, context and organisation of interactions across compartments. The unresolved question is therefore not simply which components are affected, but how their relationships collectively shape disease behaviour.

This fragmentation also creates an interpretive difficulty: apparently discordant observations may describe different aspects of the same disease process. Reduced circulating progenitor-cell numbers, for example, do not by themselves distinguish depletion of the marrow pool from defective mobilisation, increased retention or altered trafficking.[32,39,40,48–50] Similarly, evidence of vascular expansion and sprouting angiogenesis[53] coexists with reports of microvascular rarefaction and impaired marrow perfusion.[33,34,51] These findings need not be mutually exclusive. They may reflect differences in anatomical compartment, experimental model or disease stage; they do not by themselves establish a single temporal sequence. An integrated framework is needed to determine where observations converge, where their interpretation is context-dependent and where uncertainty remains.

*Study Approach and Objectives*

Addressing these interpretive challenges requires an approach that combines evidence synthesis with an explicit representation of relationships across marrow compartments. Systematic reviews help organise and critically assess published findings, providing a foundation for scientific interpretation and clinical guidance. [63],[64] Systems biology offers a complementary approach, using network reconstruction and computational modelling to examine how biological components interact across scales.[65–67] In diabetic BM disease, assembling such information is particularly challenging because the relevant evidence is dispersed across publications, experimental models, cell types and biological scales. Large language models (LLMs) may assist literature screening, information extraction and synthesis, reducing the burden of manual evidence curation.[71] However, biological reconstruction requires a clearly defined evidence base, traceable interpretation, investigator oversight and explicit assessment of the proposed relationships.

These requirements informed the design of our investigator-guided workflow, which comprised two linked stages.: (1) independent LLM interrogation of a predefined literature corpus to reconstruct diabetic BM disease, followed by quantitative graph-theoretical analysis and (2) complementary biological plausibility assessments of the frozen network. The ultimate goal was to determine whether this approach could generate a biologically coherent systems-level representation of diabetic BM dysfunction and identify regulatory relationships suitable for subsequent mechanistic investigation.

# METHODS

## Study workflow

The study analytical blocks and freeze points marking critical transitions in the workflow are illustrated in **Figure 1**,

**Block 1** reconstructed the literature-derived regulatory architecture of diabetic BM disease, including physiological relationships represented as preserved or altered in diabetes. First, a chronologically organised Discovery Corpus and a non-overlapping Validation Corpus were assembled. Both corpora were frozen before independent LLM analyses were conducted. Cross-platform comparison and investigator-guided integration then produced a biological consensus. This was translated into standardised node and edge tables encoding directed interactions. Finalisation of these tables defined the frozen reference network. The tables were then imported into Gephi for network construction and visualisation.

**Block 2** began with quantitative analysis of the frozen network and computational deletion of the endothelial node in a separate copy of the graph. Four complementary assessments followed: (A1) external biological plausibility; (A2) within-corpus physiological comparison; (A3) missing-edge sensitivity; and (A4) graph-constrained interrogation fidelity. The physiological comparison used relationships derived exclusively from the same frozen literature corpus. All computational modifications were confined to explicitly defined network variants; the frozen reference network remained unchanged. The following paragraphs illustrate the sequential analyses.

## Corpus curation and freeze

Potentially relevant publications were identified through PubMed and retrieved as full-text PDF files. The investigator selected publications for mechanistic relevance, checked for duplication and organised the corpus chronologically. The canonical register of publications and frozen corpus certification are reported in **Supplementary Files S1** and **S2**.

The *Discovery Corpus* contained 18 unique publications (2002–2025) and was designed to provide a longitudinal mechanistic framework for reconstruction. It was enriched for a coherent sequence of studies from two closely associated research programmes to which the author had contributed, together with independent studies establishing the early concepts of diabetic BM dysfunction and stem cell mobilopathy. This structure allowed progressive mechanistic observations to be examined in their chronological context but also introduced the possibility of research-lineage bias. To mitigate this risk, the emerging architecture was subsequently assessed against a broader Validation Corpus.

The *Validation Corpus* contained 43 primary studies and selected reviews or commentaries (1997–2025) drawn from the broader literature on physiological and diabetic BM biology. It was non-overlapping with the Discovery Corpus and drew predominantly on work from independent investigators. Its purpose was to assess whether the Discovery-derived architecture was supported, modified, contradicted or expanded by a wider evidence base.

## Corpus uploading

Full-text PDFs were uploaded directly to the LLM platforms without investigator-mediated text extraction, optical character recognition or reformatting. Figures, tables and legends remained within the documents, and parsing relied on each platform's native capabilities. The protocol specified chronological upload groups, with Discovery and Validation handled separately, and instructed each LLM to restrict reconstruction to the supplied publications (**Supplementary File S3**).

## AI platforms

Differences in model design and training may influence how LLMs organise and interpret the same evidence. We therefore compared reconstructions generated by three platforms—OpenAI ChatGPT, Google Gemini and DeepSeek V4—to identify shared conclusions and platform-specific interpretations. The study records identify GPT-5.6, Gemini 3.1 and DeepSeek-V4 and date the analyses to June–August 2026. During the independent reconstruction stages, each platform analysed the same frozen corpus using identical prompts in a separate analytical environment, without access to the other platforms' outputs.

## Stepwise interrogation

Analysis was initiated through investigator-issued natural-language prompts that defined the analytical task, the permitted evidential basis, the distinction between direct observation, evidence-supported inference and broader biological interpretation, and the required structure of the response. Prompt refinement was undertaken on three representative Discovery publications before definitive interrogation to improve specificity, analytical consistency and separation of evidence from interpretation. The same finalised prompt sequence was then applied across all three platforms. The complete verbatim prompts are provided in **Supplementary File S3**.

*Prompt 1 — Discovery evidence extraction*

The LLM was instructed to extract the principal experimental observations, cellular and molecular components, pathways and functional relationships from chronologically ordered groups, each encompassing three to four publications. After each group, the cumulative evidence was updated to identify findings that were confirmed, expanded or revised, newly emerging mechanisms and unresolved questions. The cycle continued until all 18 Discovery publications had been analysed.

*Prompt 2 — Discovery systems reconstruction*

Each platform was then instructed to reconstruct diabetic BM disease as an integrated biological system using only the cumulative Discovery evidence. The reconstruction addressed biological compartments, directional interactions, regulatory hierarchy, signalling organisation, functional modules, temporal evolution, biological heterogeneity, network organisation and knowledge gaps.

*Prompt 3 — Validation assessment*

Validation publications were uploaded separately, preserving chronological organisation. Each platform was asked to determine whether the major conclusions emerging from the Discovery reconstruction were supported, modified, contradicted or expanded by the Validation evidence. The platforms were asked to revise the reconstructed architecture and indicate whether confidence in the relevant conclusion increased, decreased or remained unchanged.

*Prompt 4 — Final evidence-based synthesis*

Next, each platform was asked to generate a final independent synthesis of biological organisation, regulatory hierarchy, signalling architecture, functional modules, temporal evolution, biological heterogeneity, confidence in the major conclusions and remaining knowledge gaps. Where more than one systems-level interpretation remained compatible with the evidence, the models were instructed not to force a single explanation. Each platform therefore produced its final synthesis independently before any cross-platform integration.

**Cross-platform comparison**

Following the platform-specific syntheses, the outputs were compared using the predefined concordance prompts (**Supplementary File S3**). These prompts required biological statements to be organised by compartment, interaction, pathway or systems-level interpretation, and synonymous conclusions to be merged into common rows. Stage-specific matrices recorded agreement and divergence, followed by longitudinal comparison and investigator-guided synthesis (**Supplementary Files S4–S8**). Confidence labels represented qualitative judgements of consistency and evidential support.

**Network construction, harmonisation and weighting**

Following completion of the three independent final syntheses, a separate Gemini project was used for cross-platform semantic harmonisation and data structuring. Gemini identified biologically equivalent conclusions, harmonised synonymous terminology and organised convergent and divergent interpretations into standardised propositions. The investigator reviewed the proposed consensus against the source evidence for biological plausibility, terminological accuracy, internal consistency and potential overinterpretation before network construction.

The reviewed consensus was translated into standardised node and directed edge tables through an investigator-guided procedure. Biological entities—including cell populations, molecular mediators, signalling pathways, physiological processes and disease phenotypes—were assigned consistent names. Related entities were sometimes represented within broader categories. These included haematopoietic stem cells (HSCs), haematopoietic stem and progenitor cells (HSPCs), and lineage-negative, stem cell antigen-1-positive, c-Kit-positive (LSK) cells; endothelial progenitor cells (EPCs), circulating angiogenic cells (CACs), and cluster of differentiation 34-positive ($CD34^+$) reparative-cell populations; and overlapping stromal populations. Interpretation retained the biological distinctions within these grouped categories. Nodes were annotated by biological type, compartment, functional category and proposed hierarchical role.

Relationships contained within the reviewed consensus were encoded as directed source→target interactions, with directionality reflecting the supporting evidence. Composite propositions could generate multiple edges where needed to represent distinct relationships. Harmonisation removed duplicate entries and aligned node names, interaction descriptions and directionality with the reviewed consensus. The complete node dictionary and directed edge table are provided in **Supplementary Files S9 and S10**, respectively.

Each edge retained separate agreement and confidence annotations derived from the AI-assisted consensus analysis. Agreement described support across platform outputs, whereas confidence reflected the qualitative assessment of cross-platform consistency and supporting evidence. These annotations were encoded numerically in the AgreementScore and ConfidenceScore fields. AgreementScore ranged from 1 to 3. ConfidenceScore was 4 for Very High, 3 for High or Moderate High, 2 for Moderate and 1 for Low confidence. Evidential edge weight was calculated as:

$$Weight = ConfidenceScore \times AgreementScore$$

The scheme allowed weights from 1 to 12, representing combined evidential confidence and platform agreement rather than biological interaction strength. Finalisation of the node and edge tables defined the frozen reference graph used for subsequent network analysis.

**Quantitative network analysis**

The node and edge tables were frozen and imported into Gephi (version 0.11.2; Gephi Consortium) for visualisation and quantitative analysis.[68] Within Gephi, ForceAtlas2 was used to generate the network layout,[69] and Louvain community detection was applied at resolution 1.0 to identify groups of densely interconnected nodes.[70] The principal metrics—degree, directed betweenness, weakly connected components, mean shortest-path length, global efficiency and diameter—were calculated on the unweighted graph. Evidential weights were used for weighted degree and graphical annotation. Directed betweenness excluded path endpoints and was normalised by $(N - 1)(N - 2)$, where N is the number of nodes in the analysed graph. Path length and diameter were calculated within the largest component, ignoring arrow direction. Global efficiency used the complete undirected graph, with disconnected pairs contributing zero. The same definitions were applied before and after perturbation, and results were reported descriptively.

**Targeted in silico node perturbation**

BM endothelium was selected for targeted perturbation because it emerged as the most structurally prominent cellular compartment, ranking highest among cellular nodes for connectivity and directed betweenness centrality and linking vascular, stromal, haematopoietic and signalling domains. The endothelial node and all incident edges were removed from a copy of the frozen network, and network metrics were recalculated using the same analytical procedures.

**External biological plausibility**

This analysis was performed after network freezing, using investigator-formulated questions and a defined set of curated databases. ChatGPT (OpenAI) assisted evidence retrieval and interpretation. The resources comprised the Kyoto Encyclopedia of Genes and Genomes (KEGG)[72] and connectomeDB2025,[74] complemented by Reactome[73] and Gene Ontology/Gene Ontology Causal Activity Modelling (GO-CAM)[75] for pathway and functional-process information. The investigator reviewed the supporting records and final classifications against the direction and biological meaning of each encoded interaction.

Supporting evidence and interaction-level classifications are documented in **Supplementary File S15**.

Each interaction was assessed separately. Node centrality, edge weights, deletion results and hub rankings were excluded from the classification materials. Co-occurrence alone was insufficient evidence of concordance. For composite interactions, the assessment distinguished support for individual components from support for the complete encoded relationship. The frozen reference network remained unchanged throughout the assessment.

Interactions were assigned to five mutually exclusive categories. Direct concordance indicated support for the encoded directed relationship; mechanistic concordance, support for the mechanism despite differences in context or representation; and physiological concordance, support for a broader cellular or systems-level relationship. Interactions lacking sufficient support at these levels were classified as plausible but externally unresolved. Evidence of an incompatible relationship or direction was classified as contradictory. The audit recorded qualifications concerning direction, cellular context and diabetes-specific effects alongside each classification.

After completion of the whole-network assessment, the classifications of all incoming and outgoing endothelial interactions were summarised separately because endothelium was structurally prominent in the baseline network.

**Physiological contextualisation of the frozen disease network**

The frozen diabetic BM network included relationships potentially shared with normal physiology. To provide physiological context, relationships describing normal BM regulation and non-diabetic cellular interactions were extracted exclusively from the same corpus after Freeze point 2. These were mapped to the network's standardised node definitions and compared as directed source→target pairs. Unmapped relationships were retained in the audit trail but excluded from the matched comparison (**Supplementary File S14**).

Each directed source→target pair was classified by where it appeared. Shared pairs connected the same entities in the same direction in both networks. *Physiological-reference-only pairs* appeared only in the physiological reference, whereas *unmatched diabetic-network interactions* appeared only in the diabetic reconstruction. These differences reflected incomplete reference coverage or encoding differences.

For shared pairs, we then compared the biological effect assigned to the connection. *Preserved meaning* indicated a similar regulatory role in both representations. *Functional rewiring* indicated an altered role—for example, physiological niche support represented as deficient support in diabetes—even though the same entities remained connected in the same direction.

**Targeted missing-edge sensitivity analysis**

Because literature-derived networks may omit biologically plausible relationships, we examined whether the endothelial topological phenotype was sensitive to one such missing interaction. The endothelium→stem cell factor (SCF) edge fulfilled the selection criteria for a sensitivity analysis: directional support from the physiological reference and curated evidence, unambiguous mapping to existing nodes, and absence from the frozen edge table. The edge was added to a separate graph copy, preserving the frozen 62-

node/79-edge reference. Baseline metrics were recalculated, and endothelial deletion was repeated. Because deletion also removed the added edge, this analysis assessed its effect on baseline topology and the relative magnitude of deletion-associated changes. The augmented graph is documented in **Supplementary File S13**.

**Exploratory AI-assisted interrogation-fidelity analysis**

Finally, we tested whether an AI assistant could answer questions about the reconstructed network while remaining faithful to its documented relationships. This required distinguishing recorded connections from plausible but unrecorded ones, and preserving direction and biological meaning of each interaction. The frozen node table and E001–E079 directed edge table provided the sole reference. Nine targeted questions (Q1–Q9), selected after freezing without preregistration, tested four complementary abilities. Query definitions and recorded outcomes are reported in **Supplementary File S16**.

- **Following documented routes (Q1–Q4).** These questions tested whether the assistant could trace a sequence of directed connections across biological compartments. The routes covered sympathetic/β-adrenergic control of the CXCL12–CXCR4–HSPC axis (Q1), oxidative-stress signalling through endothelium to HSPCs (Q2), S100A8/A9–RAGE–NF-κB-dependent inflammatory myelopoiesis (Q3), and adipogenic–macrophage signalling linking diabetes to systemic inflammation (Q4).
- **Rejecting an unsupported shortcut (Q5).** This question tested whether the assistant would mistakenly accept a plausible direct connection that was absent from the frozen graph. It was required to reject direct endothelium→SCF connectivity while identifying the documented indirect route. The reference for this test was the original frozen network, before the separate SCF-edge augmentation analysis.
- **Preserving biological meaning (Q6).** This question tested whether the assistant could interpret what each arrow represented along the endothelium→CXCL12→CXCR4→HSPC route. It had to retain three distinct annotations: endothelial downregulation of CXCL12, CXCL12–CXCR4 decoupling or desensitisation, and weakened CXCR4-dependent HSPC retention. The challenge was to preserve these regulatory effects when explaining the complete route.
- **Respecting direction and distinguishing direct from indirect connections (Q7–Q9).** Q7 and Q9 tested whether the assistant would reject the reversed connections CXCR4→CXCL12 and RAGE→S100A8/A9. Q8 required it to reject a direct endothelium→Akt/eNOS edge while recognising the documented indirect route through EGFR. These questions tested whether the answer followed the actual arrows and intermediate nodes in the graph.

.

# RESULTS

## Part 1. Network reconstruction

### Analysis of the Discovery Corpus identifies a concordant systems-level organisation across platforms

After harmonising biological terminology, cross-platform analysis of the Discovery Corpus identified 64 standardised biological conclusions. Of these, 49 (76.6%) were represented by all three platforms, 9 (14.1%) by exactly two and 6 (9.4%) by one platform only. Overall, at least two platforms shared 58 conclusions (90.6%). Seven of the nine two-platform conclusions were shared by ChatGPT and DeepSeek, and two by Gemini and DeepSeek. Four platform-specific conclusions came from ChatGPT and two from DeepSeek. Agreement was strongest for the underlying biological observations. Differences concerned mainly higher-order interpretation, particularly regulatory hierarchy and temporal relationships. Gemini and DeepSeek placed greater emphasis on a dominant regulatory axis, whereas ChatGPT described several interacting hubs. ChatGPT and DeepSeek more explicitly recognised parallel disease progression, while Gemini proposed a more definite temporal sequence (**Table 1** and **Supplementary File S4**).

Chronological interrogation of the Discovery Corpus generated an initial model centred on endothelial injury and reduced angiocrine support. This model linked HSPC depletion and impaired mobilisation to fewer circulating reparative cells and impaired tissue repair. Its initial assumptions included universal HSPC depletion, monotonic vascular rarefaction and hypoperfusion, and an anatomical hierarchy extending from the marrow niche through HSPCs to peripheral tissues (**Figure 2A**).

### Validation strengthens and refines the reconstructed biological model

The Validation analysis comprised 40 standardised conclusions. Harmonised assessment of the recorded platform interpretations identified three-platform agreement on 29 conclusions (72.5%). Two additional conclusions (5.0%) showed agreement between an identifiable pair, with the third platform's position unclear. Six (15.0%) showed different explicit validation judgements, while three (7.5%) remained indeterminate from the recorded summaries (**Table 2** and **Supplementary File S5**).

ChatGPT and Gemini agreed on context-dependent HSPC abundance, while ChatGPT and DeepSeek-V4 agreed on restricting the neutrophil OSM interpretation. Across all three platforms, agreement was strongest for the multicompartment organisation of diabetic BM disease, endothelial regulation, defective HSPC mobilisation, CXCL12–CXCR4-dependent trafficking, neural regulation and inflammatory myelopoiesis. Differences mainly concerned regulatory hierarchy and evidence handling. Gemini emphasised central nervous system and circadian control. ChatGPT distinguished direct diabetic evidence from contextual findings. DeepSeek-V4 emphasised niche regulation and CXCL12-dependent trafficking but classified several Discovery mechanisms as contradicted when matching Validation evidence was absent.

Validation increased mechanistic resolution. BM endothelium gained regulatory prominence through evidence of injury and protective counter-regulation, while HSPC mobilisation failure was recognised as compatible with preserved or expanded progenitor pools. Reports of vascular expansion and rarefaction supported context-dependent remodelling but did not establish a longitudinal sequence. Regulatory hierarchy was also linked more closely to functional evidence than to anatomical proximity alone (**Figure 2B**). Validation introduced or strengthened EGFR–ANGPT1 endothelial counter-regulation, S100A8/A9–RAGE inflammatory amplification, Jmjd3–H3K27me3-associated inflammatory memory, and acid sphingomyelinase–ceramide-associated migratory dysfunction (**Figure 2F**). EGFR–ANGPT1 describes a protective biological programme; a direct EGFR→ANGPT1 interaction was not encoded in the frozen graph.

Together, these findings refined the Discovery interpretation while preserving its core multicompartment organisation.

**Final Independent Syntheses converge on an integrated systems biology model**

Comparison of the Final Independent Syntheses identified 50 standardised biological conclusions. The detailed matrix recorded three-platform agreement for 36 conclusions (72%) and conceptual three-platform agreement for a further three (6%). The remaining conclusions showed partial, two-platform or platform-specific support, with some agreement patterns incompletely resolved (**Table 3** and **Supplementary File S6**). Across the three syntheses, vascular, neural, stromal, immune and haematopoietic abnormalities formed an interconnected multicompartment regulatory system. Endothelial regulation, CXCL12–CXCR4-dependent trafficking, neural control, inflammatory myelopoiesis and altered HSPC state were recurring organising elements.

Moreover, the Final Independent Syntheses provided context-dependent explanations for observations that appeared conflicting in isolation. Impaired mobilisation could coexist with preserved, proliferating or expanded marrow progenitor pools, so reduced peripheral cell availability did not necessarily indicate depletion of the marrow compartment. Reports of vascular activation, leakage and expansion were considered alongside rarefaction and hypoperfusion. Their interpretation accommodated differences in disease stage and anatomical context, without establishing a single longitudinal sequence across experimental models.

The principal disagreements concerned regulatory hierarchy and temporal ordering. ChatGPT favoured a distributed hub-and-bottleneck model without a universal initiating compartment or system-wide master regulator. Gemini assigned greater hierarchical importance to circadian regulation and proposed a more ordered neurovascular progression. DeepSeek-V4 favoured a niche-centred hierarchy with greater prominence of the marrow microenvironment and CXCL12–CXCR4-dependent regulation. The platforms also differed in the confidence assigned to selected pericyte, microRNA and other Discovery-derived pathways, and to marrow-to-peripheral-organ feedback.

The final syntheses therefore converged on the core architecture while retaining uncertainty about regulatory hierarchy, temporal sequence and selected mechanisms. These agreements and disagreements informed the subsequent consensus integration.

**Final AI-assisted consensus analysis defines the reconstructed biological architecture**

Cross-platform harmonisation by Gemini produced 48 standardised biological propositions from the three independent final syntheses. The integrated analysis recorded their inter-platform agreement, cross-stage stability and qualitative confidence . Following investigator review against the source evidence, all 48 propositions were retained in the evidence record, with supported interpretations distinguished from unresolved or unsupported claims.

The workflow linking consensus formation, network construction and subsequent analysis is summarised in **Figure 3**. The resulting biological framework integrated vascular, neural, stromal, immune and haematopoietic regulation through four complementary roles (**Figure 4**): BM endothelium as the principal compartment-level hub; CXCL12–CXCR4 as a molecular bottleneck for HSPC trafficking; S100A8/A9–RAGE as an inflammatory amplification pathway; and HSPC state and marrow-to-blood trafficking as the functional point of convergence.

Convergence was strongest for multicompartment organisation, endothelial involvement, mobilopathy, sympathetic–circadian control and inflammatory myelopoiesis. The initiating compartment, precise temporal sequence, complete marrow–organ feedback loop and existence of a universal master regulator remained unresolved. **Supplementary File S7** presents the integrated framework and confidence assignments, while **Supplementary File S8** traces how the platform interpretations developed across analytical stages and where disagreements persisted.

**Consensus reconstruction generates a directed multicompartment network**

Investigator-guided structuring of the reviewed consensus produced standardised tables containing 62 biological entities and 79 directed regulatory interactions (**Supplementary Files S9** and **S10**). Their finalisation defined the frozen reference network. The tables were imported into Gephi for network construction and visualisation, completing the transition from biological synthesis to quantitative network representation.

The network integrated systemic metabolic stress with vascular, neural, stromal, inflammatory and haematopoietic processes. Each directed interaction represented an encoded relationship between biological entities or processes and retained separate agreement, confidence and evidential-weight annotations. Entities without encoded connections were retained as part of the predefined consensus model. The resulting reference graph provided the basis for the analyses reported in Part 2.

**Part 2. Network analysis**

**Table 4** summarises the analyses performed on the frozen reference network and its computationally modified variants.

Part 2 began with quantitative analysis of the network constructed in Gephi. Mean out-degree—the average number of outgoing interactions per node—was 1.274. Mean shortest-path length was 2.667, and network diameter was 7, calculated using directed paths between reachable node pairs within the largest weakly connected component (**Figure 5A**; **Supplementary Screenshot S11**). BM endothelium emerged as a prominent integrative node linking vascular, inflammatory and haematopoietic processes. It ranked first by total degree, with 17 interactions, and directed betweenness centrality (**Figure 5B**). To examine its contribution to network organisation, we computationally removed the endothelial node and its incident interactions from a separate copy of the frozen graph.

After deletion, the network contained 61 nodes and 62 directed interactions. The number of weakly connected components increased from 18 to 19, and network diameter increased from 7 to 8 (+14.3%). Mean shortest-path length remained essentially unchanged, increasing from 2.667 to 2.676 (**Figure 5C**; **Table 5**; **Supplementary**

**Screenshot S12**). Thus, endothelial removal produced additional fragmentation and increased the longest shortest path, while the average distance between reachable nodes changed little. This average describes the pairs that remained reachable and does not imply preservation of overall connectivity.

Together, these findings support an integrative role for endothelium within the reconstructed network. Its removal altered network organisation and reduced connectivity. These structural effects identify relationships for biological investigation without establishing the consequences of endothelial dysfunction in vivo.

**External biological plausibility supports the reconstructed regulatory backbone**

The external plausibility audit classified 60 of the 79 reconstructed interactions (75.9%) as positively concordant. Direct concordance accounted for 18 interactions (22.8%), mechanistic concordance for 34 (43.0%) and physiological concordance for 8 (10.1%) (**Figure 6A**). The remaining 19 interactions (24.1%) were classified as biologically plausible but externally unresolved; none were classified as contradictory. Within the endothelial neighbourhood, 15 of 17 incident interactions (88.2%) showed positive concordance: 3 direct, 9 mechanistic and 3 physiological. Two remained unresolved (**Supplementary File S13**).

Supporting records covered CXCL12–CXCR4 signalling, SCF-dependent haematopoietic regulation, endothelial and angiocrine support, S100A8/A9–RAGE–NF-κB inflammatory signalling, β-adrenergic niche regulation and Akt/endothelial nitric oxide synthase (eNOS) signalling. Evidence limitations concerned the precise direction, cellular context or diabetes-specific meaning of individual relationships, particularly those represented by composite nodes. These qualifications also applied to some positively classified interactions. The findings therefore support the plausibility of the underlying regulatory systems, while leaving selected disease-state effects unresolved.

**Within-corpus physiological comparison identifies shared and unmatched relationships**

To place the diabetic network in physiological context, we compared its directed connections with relationships describing normal BM regulation within the same frozen corpus. The physiological reference contained 33 relationships. Of these, 32 could be compared with the diabetic network and involved 24 of its 62 nodes. One relationship involving a central nervous system/hypothalamic clock source was excluded because that entity was absent from the diabetic network (**Supplementary File S14**). The other 38 diabetic-network nodes mainly represented inflammatory and stress-response pathways, epigenetic regulation, vascular remodelling, systemic disease modifiers and therapeutic interventions. The physiological reference therefore covered only part of normal BM biology.

Of the 32 comparable relationships, 14 (43.8%) connected the same entities in the same direction in both networks, whereas 18 (56.3%) appeared only in the physiological reference. Conversely, 65 of the diabetic network's 79 interactions had no matched physiological counterpart. Excluding five therapeutic interactions—two involving plerixafor/CXCR4 antagonism and three involving metabolic or antioxidant rescue—left 60 unmatched non-intervention interactions, representing 75.9% of the diabetic network (**Table 6**).

These 60 interactions spanned nine functional categories, dominated by inflammatory myelopoiesis and the endothelial–vascular niche. Other categories included persistent epigenetic/transcriptional programmes and altered neural and HSPC regulation (Table 7). These descriptive categories were distinct from graph-derived communities.

We then asked whether connections shared by both networks retained the same biological meaning. Several connected the same entities in the same direction but carried different regulatory interpretations (**Table 8**). Endothelial regulation of CXCL12 was annotated as downregulation in diabetes, while CXCL12→CXCR4 signalling was described as decoupled or desensitised. SCF→HSPC and Notch→HSPC relationships were likewise represented as deficient maintenance or support. Functional rewiring therefore referred to changes in the meaning assigned to a connection, even when its direction was preserved.

Other shared connections retained similar interpretations. Akt/eNOS remained supportive of endothelial function, sympathetic activity remained linked to β-adrenergic signalling, and sensory nerves retained substance P- and calcitonin gene-related peptide (CGRP)-associated signalling. Neural regulation of CXCL12 persisted within a pathological mobilisation context. Marrow adipocyte→HSPC inhibition was also retained, with a more explicit lipotoxic interpretation. Thus, shared connectivity encompassed both functional continuity and altered regulatory effects.

To identify which aspects of normal niche support were incompletely represented in the diabetic graph, we examined the 18 physiological-only relationships. These included CXCL12 production by perivascular stromal cells and Nestin$^{+}$/leptin receptor-positive (LepR$^{+}$)/CXCL12-abundant reticular (CAR)-like stromal subsets, together with stromal and endothelial contributions to SCF. Other relationships described direct stromal and pericyte support of HSPCs, CXCR4-dependent retention, neuropeptide-mediated mobilisation and generation of circulating reparative cells. Collectively, they outlined a distributed homeostatic support system.

Finally, we compared how extensively key regulators were connected and how often they linked other nodes through the shortest directed routes. In the matched physiological subnetwork, HSPCs had the most direct connections (11), while CXCL12 and endothelium each had six. Endothelium ranked fifth by directed betweenness, a measure of how frequently a node lies on these connecting routes. In the diabetic network, endothelium had both the most direct connections (17) and the highest directed betweenness. It therefore occupied a more prominent connecting position in the diabetic reconstruction. However, the physiological reference covered fewer biological entities and relationships. The difference in endothelial ranking could consequently reflect differences in network construction as well as biology, and could not be attributed solely to diabetes

Overall, shared connections retained or changed their regulatory meaning. Because the physiological reference was incomplete, unmatched relationships could not establish diabetes specificity or biological loss. Five of the six physiological endothelial relationships were shared; the absent endothelium→SCF edge prompted a sensitivity analysis examining how its addition affected baseline topology and the effects of endothelial deletion.

**Missing-edge sensitivity shows limited baseline effects of adding endothelium→SCF**

Adding the physiological endothelium→SCF relationship to a separate augmented graph produced only minor changes in baseline topology. Mean shortest-path length decreased from 2.779 to 2.754, and global efficiency increased from 0.218 to 0.220. The largest component, number of weakly connected components and diameter were unchanged (**Figure 6B; Supplementary File S15**).

Endothelial deletion from the augmented graph produced the same final structure as deletion from the frozen graph: 19 weakly connected components, a 43-node largest component, mean shortest-path length of 3.792, diameter of 10 and global efficiency of 0.170 (**Table 9**). Relative to the augmented baseline, deletion increased mean shortest-path length by 37.7% and reduced global efficiency by 22.9%.

Both graphs became identical after endothelial deletion because the added edge was also removed. The larger relative changes reflected the modestly improved augmented baseline. This test assessed one specific omission and did not establish robustness to missing edges more generally.

**Graph-constrained interrogation preserves route structure, edge semantics and directionality**

Finally, we tested whether an AI assistant could answer questions about the reconstructed network without inventing connections or misinterpreting their meaning. We submitted 9 targeted questions and compared the answers with the relationships encoded in the frozen edge table. **Supplementary File S16** summarises the queries, expected graph relationships and recorded outcomes.

The assistant recovered documented routes passing through several intermediate nodes (Q1–Q4) and rejected an unsupported direct connection while identifying a valid indirect route (Q5). It preserved the distinct biological effects encoded along a signalling route (Q6), rejected connections presented in the reverse direction (Q7 and Q9), and distinguished an absent direct connection from an available indirect route (Q8).

All nine questions were answered correctly within this selected test set (**Figure 6C**; **Table 10**). These findings support faithful interpretation of the encoded network.

# Discussion

Original and review articles have provided valuable insights into the anatomical and functional alterations of BM niches in diabetes. This study demonstrates the feasibility of integrating literature-derived biological relationships into a regulatory network of diabetic BM disease. Human–AI collaboration contributed through complementary forms of reasoning. LLMs compared dispersed descriptions and organised candidate relationships in a common format. The investigator assessed experimental context and distinguished documented interactions from interpretative links. Comparing independently generated reconstructions suggested connections across compartments and reconciled discrepancies from the literature. Topological analysis then made proposed dependencies explicit and measurable. Structural tests and external plausibility assessment examined different aspects of the model. The detailed documentation of the evidence base, prompts, node and edge tables, and analytical procedures supports independent scrutiny and reproduction of the analyses. It also enables other investigators to extend and refine the network as new evidence emerges.

**From network structure to mechanistic understanding**

Complementary analyses identified BM endothelium as the principal integrative node in the reconstructed network. It had the highest number of direct connections and ranked first in directed betweenness centrality, indicating a prominent position along routes linking other biological entities. Removing the endothelial node increased fragmentation and network diameter, meaning that more steps were needed to connect the most distant reachable nodes. When connections were analysed without considering their direction, removal also increased average path length and reduced global efficiency. The external plausibility audit classified most endothelial interactions as concordant at different levels of support, while retaining qualifications about selected disease-state effects. Together, these findings prioritise endothelial coordination as a mechanistic hypothesis. As a note of caution, topological prominence does not necessarily confer biological or pathogenic causality.

Pericytes and perivascular stromal cells provide the immediate cellular context for endothelial regulation.[13,42] Within the functionally annotated endothelial–vascular niche module, the pericyte→endothelium relationship was interpreted as changing from vascular support in the physiological reference to detachment or dysfunction in diabetes. Altered stromal trophic interactions added another level of influence on endothelial behaviour. Principal angiocrine relationships also showed functional rewiring: endothelial support of HSPCs became deficient, endothelial→Notch signalling was represented as suppressed, and Notch→HSPC and SCF→HSPC relationships conveyed impaired maintenance. Alongside Src/Pyk2–VE-cadherin-associated barrier dysfunction, these changes suggest that vascular injury can compromise both the physical organisation and trophic activity of the niche.[11,13,35]

The HSPC maintenance and mobilisation module links niche disturbances to cell localisation and function. The frozen network represented endothelial CXCL12 downregulation and annotated CXCL12→CXCR4 signalling as decoupled or desensitised. These annotations describe altered regulatory effects while preserving the direction of the connections. Their interpretation nevertheless depends on the cellular source of CXCL12 and the experimental context; they should not imply uniformly weakened retention throughout the marrow. Preserved progenitor populations can coexist with impaired mobilisation, reflecting disruption of the balance between niche support, retention and release. Evidence that niche dysfunction impairs mobilisation and that CXCR4 antagonism can bypass some mobilisation defects supports this distinction between progenitor abundance and trafficking competence.[40],[49]

The biologically annotated neural–circadian module converges on the same trafficking machinery. Sympathetic β-adrenergic signalling and Per2/Bmal1-associated clock regulation influence CXCL12, while substance P/NK1R and CGRP-associated pathways connect sensory inputs to progenitor and niche function. These sensory effects can include direct actions on progenitor cells, without requiring an endothelial intermediary.[16–19],[36–38],[54–56] Several neural connections were retained in the physiological comparison, although their regulatory meaning was altered in the diabetic reconstruction. Together, these relationships place trafficking under coordinated neural and local niche control, helping explain how preserved connections can accompany dysfunctional cell mobilisation.

The inflammatory myelopoiesis module provides a complementary interpretation of altered cell output. S100A8/A9–RAGE-associated signalling links hyperglycaemia to increased myelopoiesis,[58] while macrophage-associated OSM connects inflammation, niche regulation and mobilopathy.[39],[43] These relationships suggest how myeloid expansion and defective mobilisation can coexist without uniform HSPC depletion. Their interpretation must retain cellular context: a separate study found neutrophil-derived OSM dispensable for HSPC trafficking, arguing against a universal OSM-dependent mechanism.[47] These inflammatory mechanisms intersect with CXCL12 regulation and other niche inputs, providing candidates for selective perturbation rather than a single obligatory pathway.

Sphingolipid and adipogenic remodelling offer additional links between metabolic stress and niche dysfunction. Acid sphingomyelinase–ceramide biology connects diabetes to an altered balance of reparative and inflammatory marrow cells,[60] while RhoA/ROCK signalling highlights a mechanism of endothelial barrier dysfunction.[35] The reconstruction brings these processes into relation with oxidative stress and supportive Akt/eNOS signalling. Adipocyte-associated inhibition and an MCP-1 feedback loop between adipocytes and stromal cells add potential routes through which metabolic disturbance can affect HSPCs and their niche.[20],[45] These linked processes suggest that restoring cell availability may require attention to the supporting environment as well as the progenitor cells themselves.

In this respect, the possibility that epigenetic memory sustains dysfunction after the initiating stress has subsided adds a temporal dimension to this systems biology interpretation. Jmjd3-dependent changes in H3K27 methylation have been linked to inflammatory programming of marrow progenitors and their progeny in diabetes.[59] The miR-155–FOXO3a pathway provides another connection between diabetic niche injury and progenitor-cell dysfunction.[34] Evidence for miR-21/PDCD4-mediated injury comes from circulating CD34+ cells and endothelial cells in critical limb ischaemia, and should not be treated as direct verification of an entire marrow-specific TUG1–miR-21–PDCD4 pathway.[46] The composite non-coding RNA relationships in the reconstructed model therefore remain candidates for more specific testing. Perturbation and rescue experiments could distinguish defects corrected by restoring endothelial–perivascular support from those maintained by persistent inflammatory or epigenetic programmes, identifying where combined interventions may be needed.

Preserved connections with altered regulatory effects also prompt an allostatic interpretation of diabetic mobilopathy. We hypothesise that, alongside direct injury, changes in HSPC activity and trafficking may partly reflect adjustments that favour maintenance of the resident cell pool under sustained systemic stress. Altered metabolism and energy allocation could contribute to this balance, potentially preserving local viability at the expense of BM responsiveness and reparative capacity. Epigenetic memory might stabilise these responses, limiting recovery after metabolic improvement. This interpretation extends beyond what the static reconstruction can establish. Longitudinal measurements of metabolism, epigenetic state, niche integrity and HSPC behaviour would help distinguish compensatory adaptation from persistent or progressive dysfunction.

**Challenging the AI-assisted reconstruction**

Assessment against external curated resources is an important component of evaluating the biological plausibility of literature-derived networks. It becomes particularly valuable in investigator-led AI reconstruction of complex phenomena, where coherent interpretations and agreement between platforms may still contain shared errors. Block 2 therefore challenged the architecture generated in Block 1 through complementary structural, biological and interrogation assessments. The reference network remained frozen, allowing its interpretation to be examined without changing its content to accommodate subsequent findings.

Centrality and endothelial deletion supported an integrative role for endothelium within the reconstructed network. Their contribution was to identify structural dependencies for biological testing, while causal hierarchy remained unresolved.

Concordance with KEGG and complementary resources supported the biological basis of many encoded relationships. Direct concordance supported a specified interaction, whereas mechanistic and physiological concordance established compatibility at broader levels. The physiological comparison further showed how preserved connections could carry different regulatory meanings. Together, these assessments refined the mechanistic interpretation above by distinguishing functional continuity, context-dependent rewiring and uncertainty in disease-specific annotations.

Missing-edge sensitivity analysis examined the influence of incomplete encoding. Adding the omitted, but plausible endothelium→SCF relationship modestly altered baseline connectivity and the relative effects of endothelial deletion. As new research identifies additional interactions, incorporating them into updated network versions could improve biological coverage, reveal previously unrecognised dependencies and strengthen the model's value for guiding experimental investigation.

Finally, interrogation tests showed that an AI assistant could preserve documented routes, directionality and biological meaning while rejecting unsupported connections within the selected query set. This supports the practical use of the reconstruction for exploring encoded relationships and developing mechanistic questions. Collectively, these assessments refined the interpretation of the initial consensus and strengthened its value as a testable systems model, while identifying where further evidence is needed.

## Conclusions and perspectives

The present work may stimulate new computational, experimental, and translational studies of BM dysfunction in diabetes. For instance, one could investigate which inferred regulatory dependencies can be targeted to restore BM function in diabetes. Perturbation and rescue experiments should assess recovery of niche support, HSPC trafficking and reparative function, alongside cell abundance, and define the appropriate direction, timing and duration of intervention. Spatial and single-cell measurements could identify the cellular sources of relevant signals, while longitudinal studies could distinguish reversible changes from persistent dysfunction. Together, these data could support dynamic modelling of disease progression and recovery using logical or stochastic platforms such as Cell Collective, GINsim or MaBoSS.[76–78] This transition requires experimentally supported regulatory rules and defined initial biological states, beyond the structural information contained in the present network. In parallel, patient studies could determine whether combined measures of niche function and haematopoietic state identify clinically meaningful patterns of diabetic BM dysfunction. These advances would connect systems-level hypotheses to functional biomarkers and help prioritise interventions that restore vascular support, improve trafficking or limit inflammatory persistence, with their value ultimately determined by biological and clinical benefit.

# FIGURES

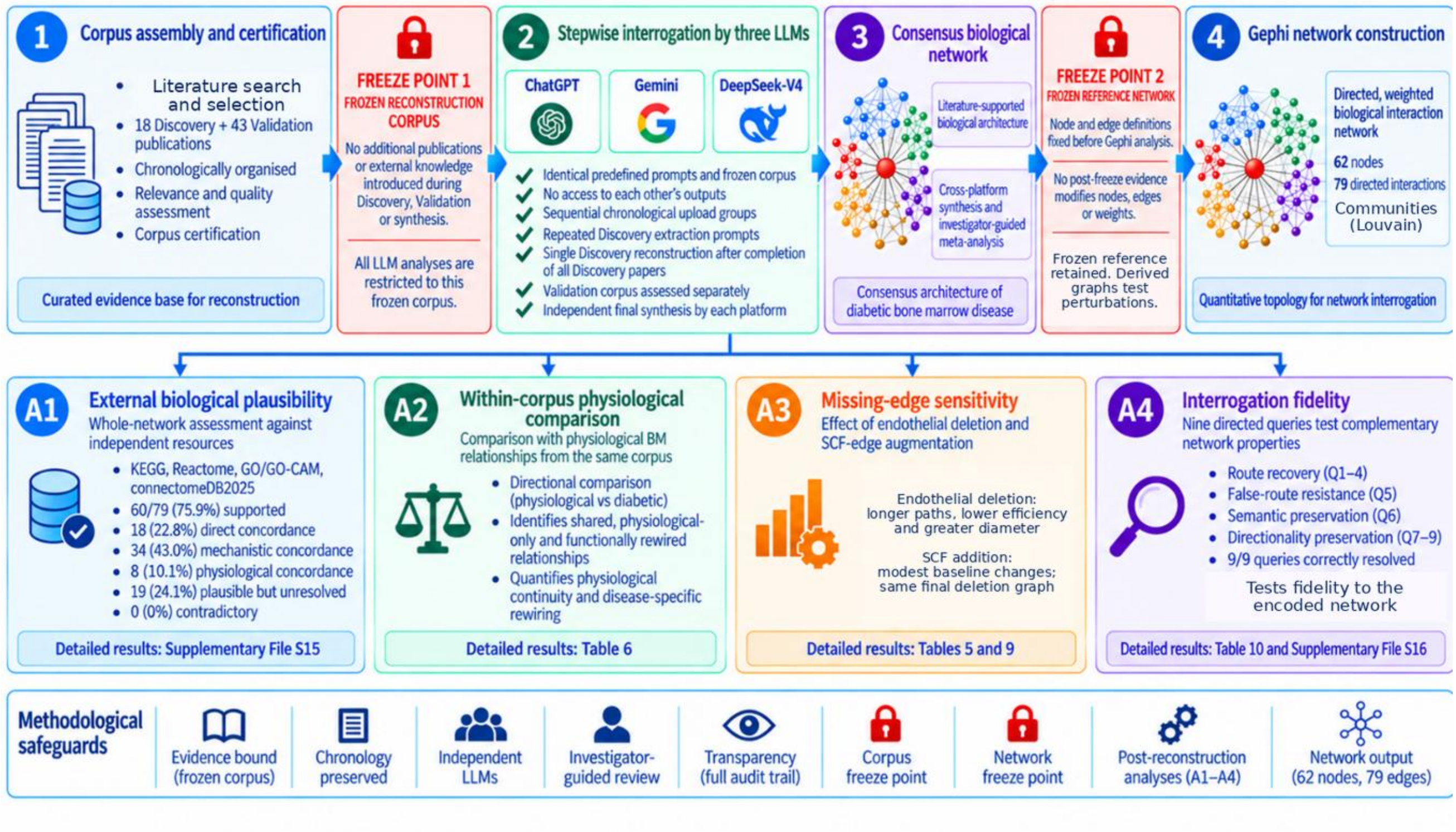

**Figure 1. Reconstruction and assessment of the diabetic bone marrow network.** (1–4) Corpus assembly, separate interrogation by three large language models, consensus integration and Gephi network construction. Freeze point 1 fixes the literature corpus; freeze point 2 fixes the final node and edge tables before Gephi import. (A1–A4) External plausibility, within-corpus physiological comparison, missing-edge sensitivity and interrogation fidelity examine the reconstruction. Perturbations use derived graphs; the 62-node, 79-interaction reference network remains unchanged.

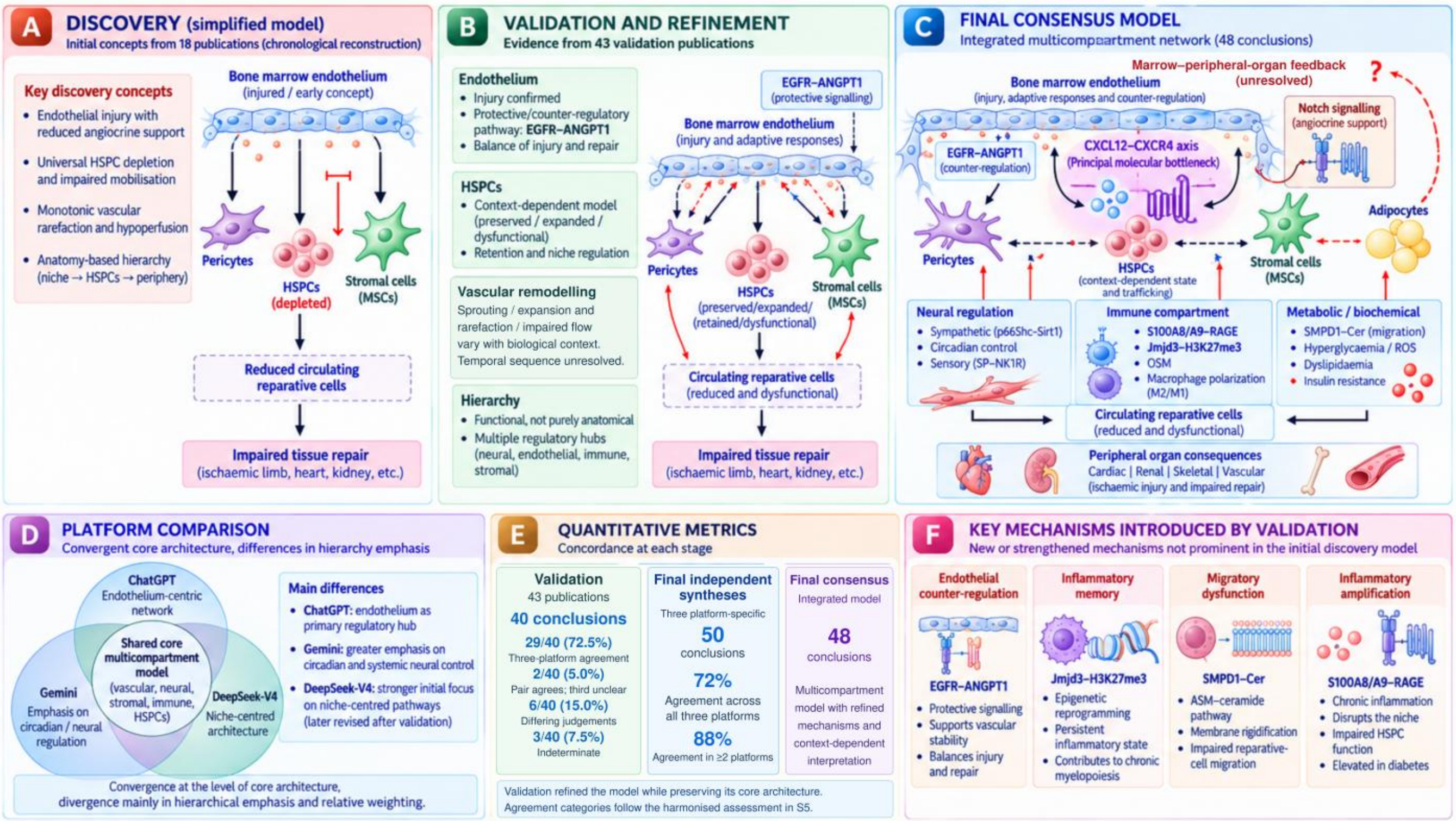

**Figure 2. Evolution from discovery to final consensus.** (A) Initial model based on 18 Discovery publications. (B) Assessment against 43 Validation publications introduced endothelial counter-regulation and context-dependent progenitor and vascular responses. The temporal relationship between vascular expansion and rarefaction remains unresolved. (C) The final model integrates 48 conclusions across interacting marrow compartments; peripheral-organ feedback remains unresolved. (D) Platform outputs converge on the multicompartment architecture but differ in regulatory emphasis. (E) Among 40 Validation conclusions, 29 (72.5%) showed three-platform agreement, two (5.0%) agreement between an identifiable pair with the third position unclear, six (15.0%) differing judgements and three (7.5%) indeterminate agreement. Independent syntheses yielded 50 conclusions, with 72% agreement across all three platforms and 88% across at least two. (F) Validation strengthened the EGFR–ANGPT1 endothelial counter-regulatory programme, Jmjd3–H3K27me3-associated inflammatory memory, SMPD1–ceramide-associated migratory dysfunction and S100A8/A9–RAGE inflammatory amplification.

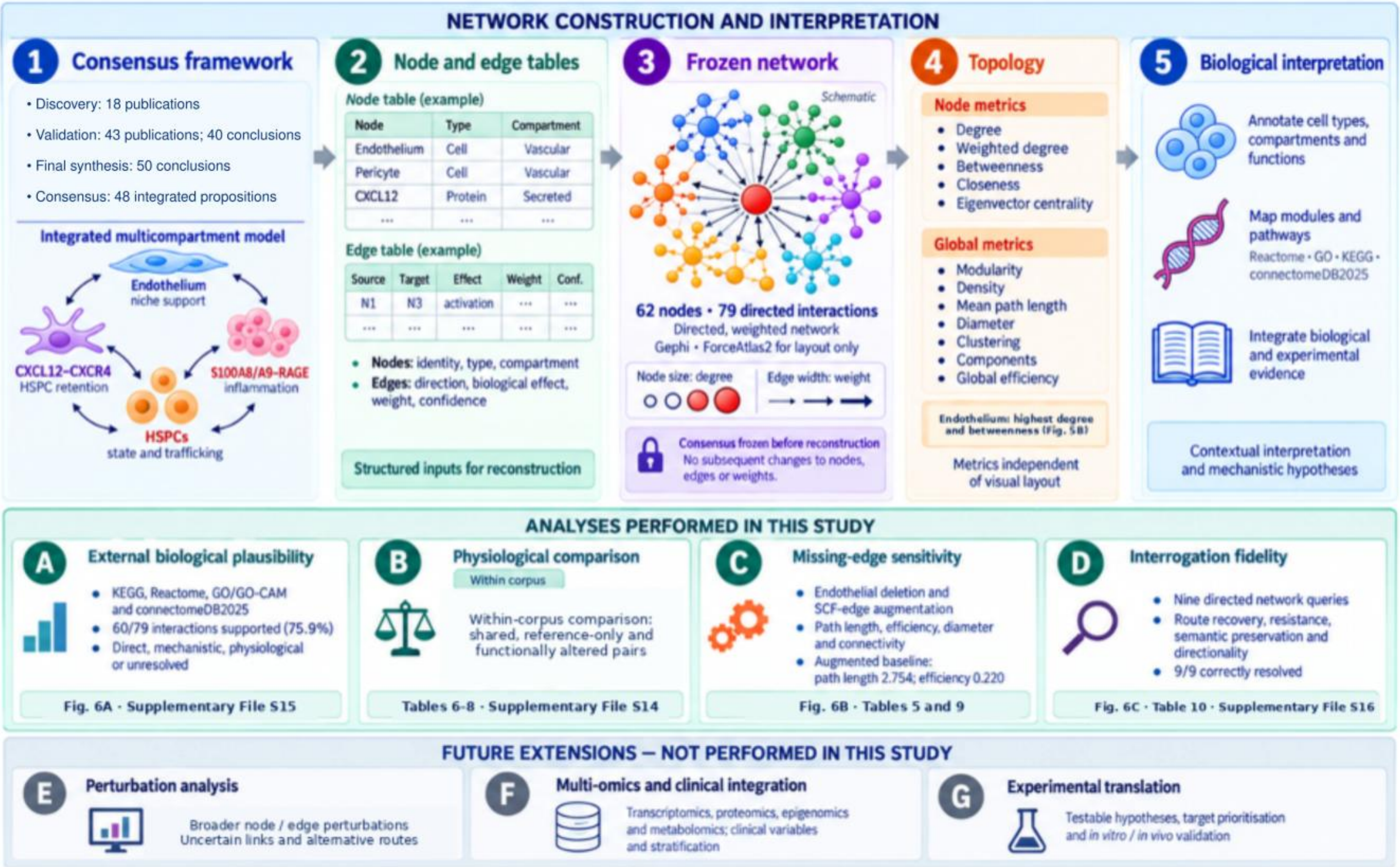

**Figure 3. From evidence synthesis to network interpretation.** (1–5) Consensus biology is encoded in node and edge tables, reconstructed as a directed network in Gephi, characterised topologically and interpreted biologically. The example tables and network drawing are schematic. Node size denotes degree; edge width denotes evidential weight, not biological interaction strength. ForceAtlas2 determines display positions, not network metrics. (A–D) Analyses performed here assess external plausibility, physiological relationships, missing-edge sensitivity and interrogation fidelity. (E–G) Broader perturbation studies, multi-omics and clinical integration, and experimental translation are proposed extensions. Subsequent assessment preserves the frozen reference; perturbations examine explicitly defined variants.

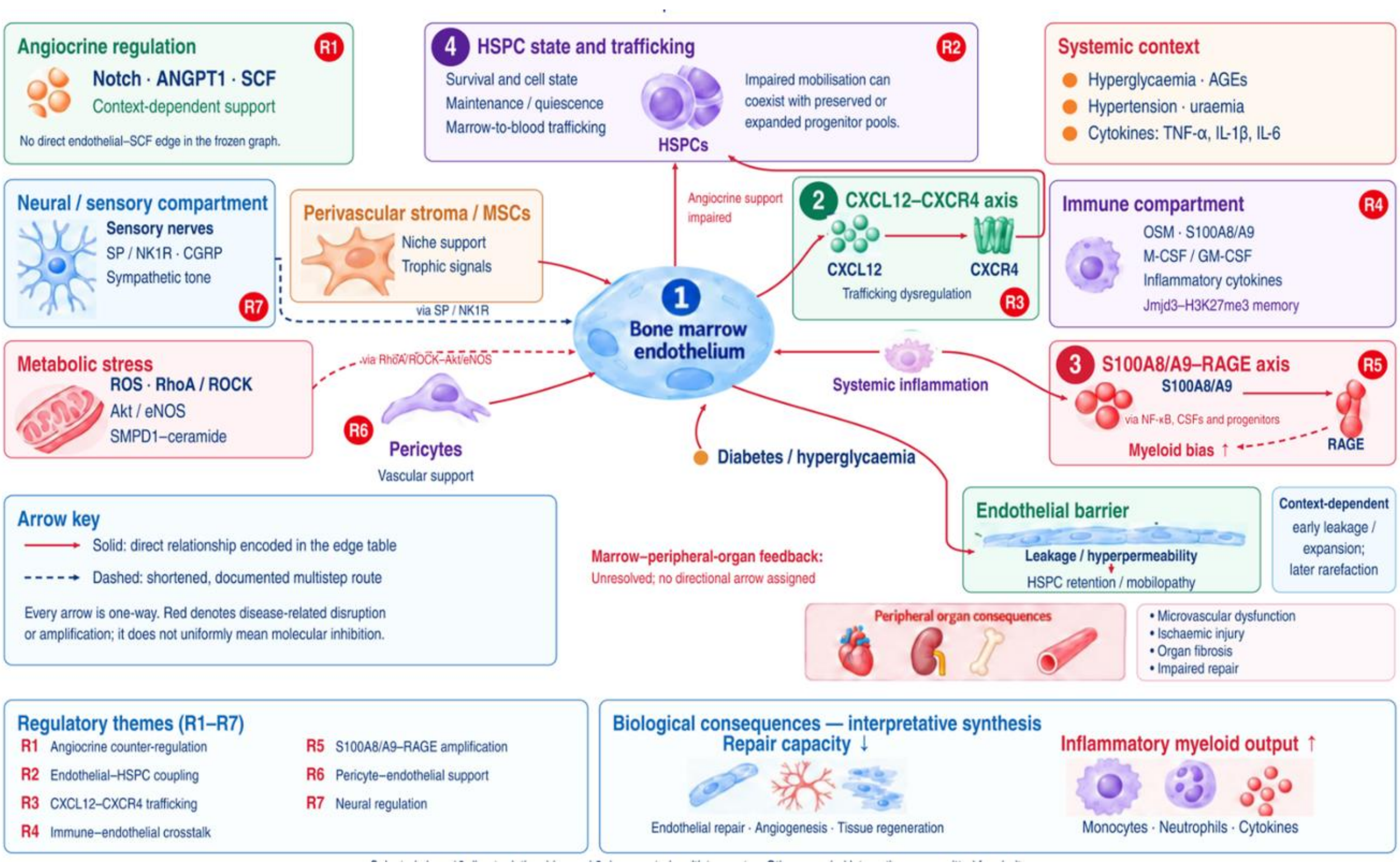

**Figure 4. Interacting regulatory domains in diabetic bone marrow disease.** (1) Bone marrow endothelium, (2) CXCL12–CXCR4 signalling, (3) S100A8/A9–RAGE inflammatory amplification and (4) haematopoietic stem and progenitor cell (HSPC) state and trafficking. Solid arrows show selected encoded direct relationships; dashed arrows summarise documented multistep routes. Arrowheads indicate direction. Red denotes disease-related disruption or amplification, not uniform inhibition. R1–R7 label regulatory themes, not individual edge identifiers. Biological consequences are interpretative summaries; marrow–peripheral-organ feedback remains unresolved. MSCs, mesenchymal stromal cells; SCF, stem cell factor.

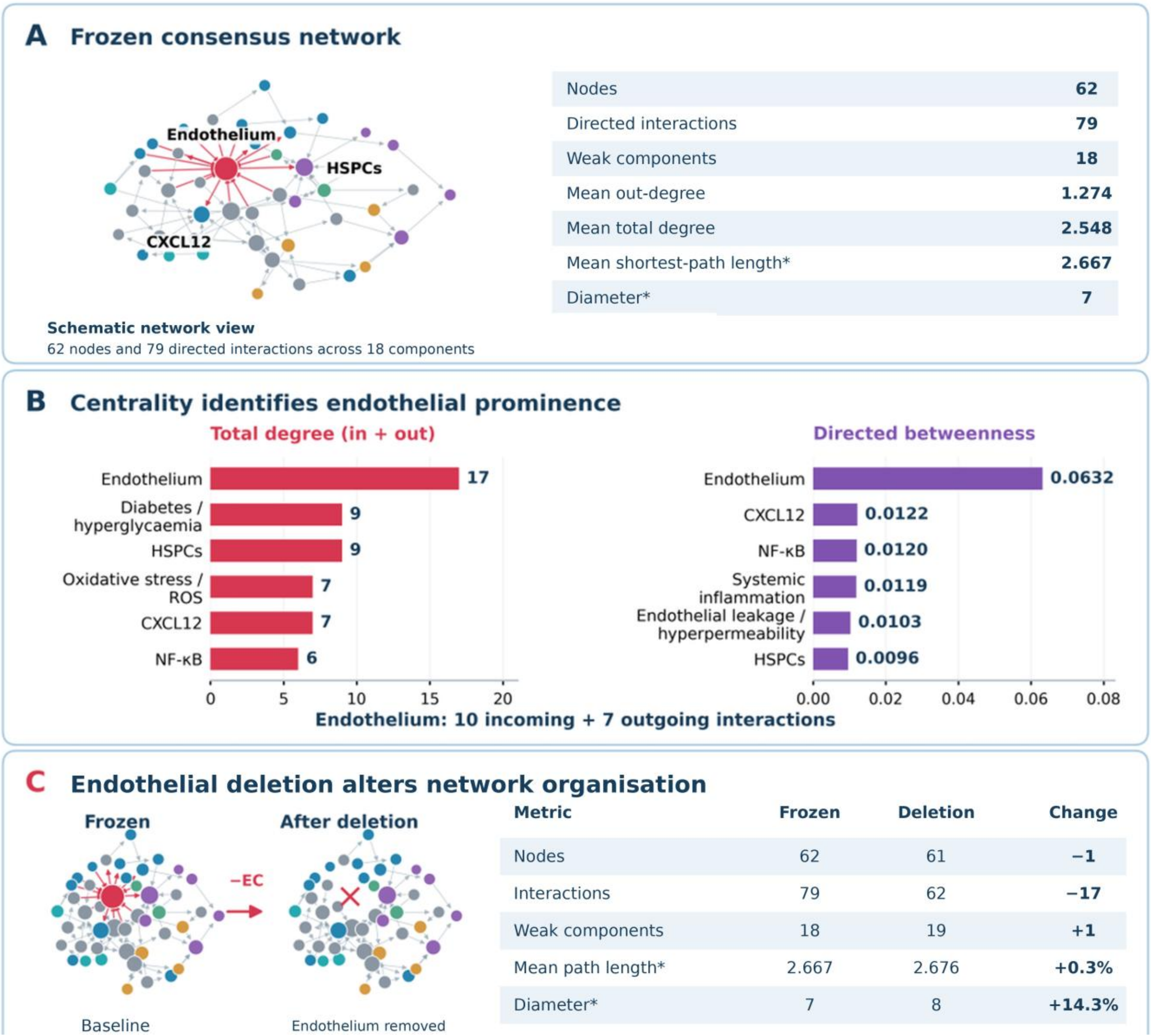

**Figure 5. Network organisation and endothelial integration.** (A) The frozen network contains 62 nodes and 79 directed interactions; 17 isolated nodes are shown separately. Node size denotes total degree; colours denote biological groups, with endothelium and its incident edges highlighted in red. (B) Endothelium ranks first by unweighted total degree and directed betweenness. (C) Endothelial deletion increases fragmentation and path length and reduces efficiency. Node positions are fixed across conditions. Path length and diameter use the largest undirected component; efficiency uses the full undirected graph. Path length and diameter use unweighted directed paths between reachable node pairs within the largest weakly connected component. These changes measure structural dependence, not biological causality.

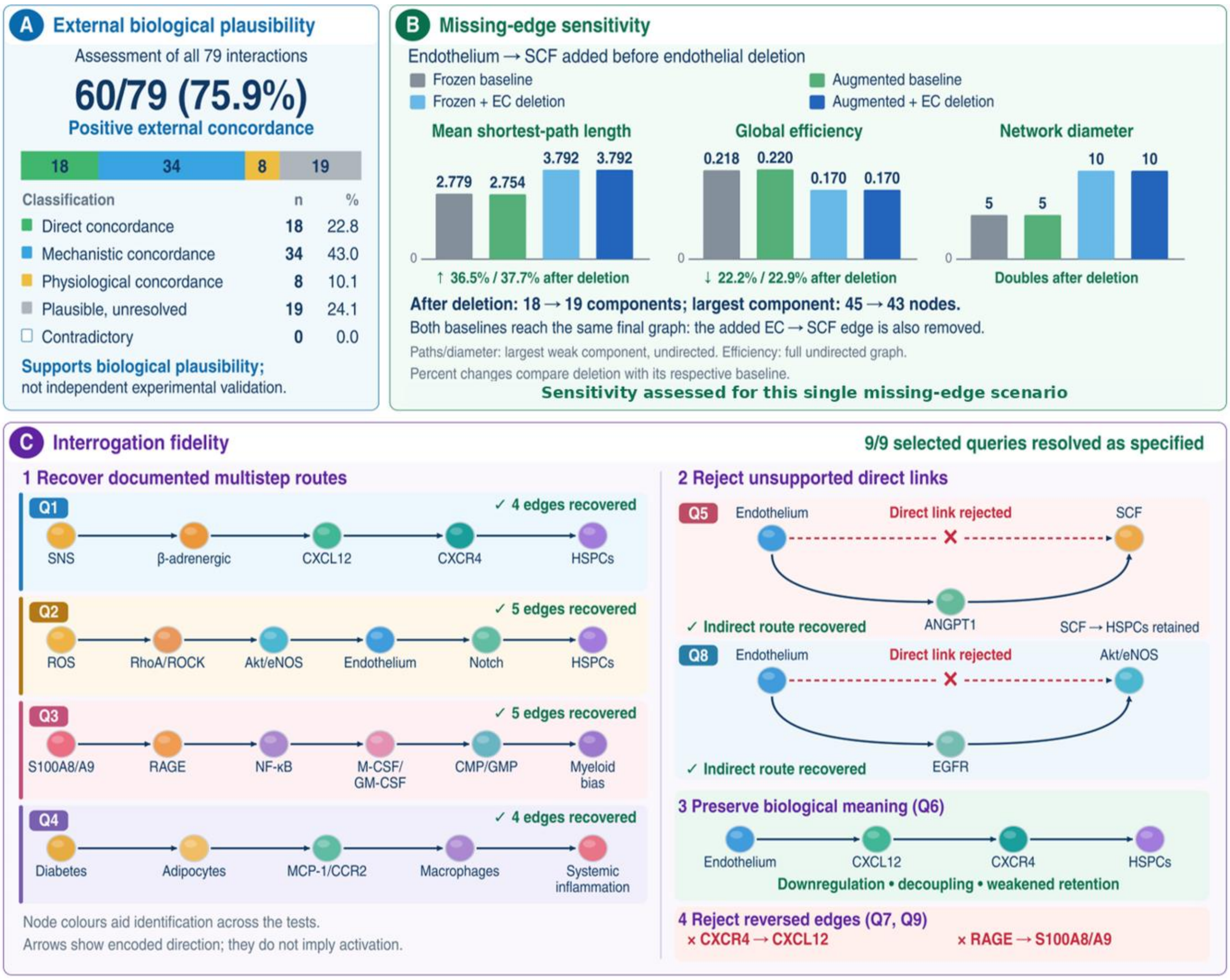

**Figure 6. Complementary assessments of the frozen network.** (A) External resources support 60/79 interactions (75.9%) at direct, mechanistic or physiological levels; 19 remain unresolved and none are classified as contradictory. (B) Adding endothelium→stem cell factor (SCF) modestly changes baseline topology. Deletion then yields the same final graph because the added edge is removed with endothelium. Path length and diameter use the largest undirected component; efficiency uses the full undirected graph. (C) All nine selected queries recover the required routes or reject unsupported connections while preserving the tested meanings and directions. Node colours aid identification; arrows do not imply uniform activation. Concordance and query fidelity do not independently establish biological validity.

# TABLES

**Table 1. Discovery concordance**

| Classification | n/N | % |
|---|---|---|
| Reproduced by all three platforms | 49/64 | 76.6 |
| Reproduced by at least two platforms | 58/64 | 90.6 |
| Platform-specific | 6/64 | 9.4 |
| Total standardised conclusions | 64/64 | 100 |

**Legend.** Cross-platform concordance of biological conclusions generated independently by ChatGPT, Gemini and DeepSeek-V4 during interrogation of the Discovery Corpus. Concordance was assessed after harmonisation of synonymous terminology. The "at least two" category includes conclusions recovered by all three platforms.

**Table 2. Harmonised cross-platform agreement in the Validation analysis**

| Agreement category | Conclusions, n (%) | Platform attribution |
|---|---|---|
| Three-platform agreement on the conclusion or its revision | 29 (72.5%) | ChatGPT, Gemini and DeepSeek-V4 |
| Identifiable agreeing pair; third position unclear | 2 (5.0%) | ChatGPT–Gemini: 1; ChatGPT–DeepSeek-V4: 1 |
| Different explicit validation judgements | 6 (15.0%) | ChatGPT and DeepSeek-V4 differed; Gemini's position was unclear |
| Indeterminate agreement pattern | 3 (7.5%) | Insufficient detail for reliable attribution |
| Total | 40 (100.0%) | |

**Legend.** Agreement was assessed retrospectively from the platform interpretations recorded in Supplementary File S5. Compatible biological judgements could coexist with differences in classification or confidence. An unclear position was treated as insufficient information rather than disagreement. The two pair-level conclusions therefore do not establish agreement by exactly two platforms.

**Table 3. Cross-platform concordance of the Final Independent Syntheses.**

| Concordance category | Conclusions, n | Percentage |
|---|---|---|
| Explicit three-platform agreement | 36 | 72% |
| Conceptual three-platform agreement | 3 | 6% |
| Two-platform or two-to-three-platform agreement | 6 | 12% |
| Explicit support from one platform | 4 | 8% |
| Uncertain one-to-two-platform support | 1 | 2% |
| **Total** | **50** | **100%** |

Categories reproduce the classifications recorded in the detailed matrix of Supplementary File S6. Conceptual agreement includes partially expressed support for the shared biological interpretation. Mixed categories retain uncertainty about the extent of platform agreement. Explicit support from one platform does not establish disagreement by the others. Percentages describe concordance between recorded interpretations, rather than biological accuracy.

**Table 4. Overview of post-reconstruction analyses**

| Analytical question | Approach | Headline finding |
|---|---|---|
| How is the network organised? | Directed topology and centrality | Bone marrow endothelium ranked first by degree and betweenness |
| Is endothelial prominence structurally consequential? | Endothelial deletion | Path length increased by 36.5% and global efficiency decreased by 22.2% |
| Are reconstructed interactions biologically plausible? | KEGG, connectomeDB2025, Reactome and GO/GO-CAM interrogation | 60/79 interactions showed concordance at direct, mechanistic or physiological level |
| Which relationships are shared with the partial physiological reference? | Within-corpus directed-pair comparison | 14 shared pairs; 60 unmatched non-intervention interactions |
| Are retained relationships functionally unchanged? | Comparison of edge semantics | Several homeostatic relationships were retained structurally but functionally rewired |
| How does a specified missing edge affect the baseline and deletion comparison? | Endothelium→SCF augmentation | Small baseline changes; identical final graph because the added edge was deleted |
| Can the frozen graph be interrogated faithfully? | Nine graph-constrained queries | Valid paths were recovered and unsupported or reversed relationships rejected |

**Legend.** Summary of the principal post-reconstruction analyses applied to the frozen 62-node/79-edge network, the analytical question addressed by each approach and the corresponding headline result.

**Table 5. Effect of bone marrow endothelial deletion on network topology**

| Metric | Frozen network | After endothelial deletion | Change |
|---|---|---|---|
| Nodes | 62 | 61 | −1 |
| Directed interactions | 79 | 62 | −17 |
| Connected components | 18 | 19 | +1 |
| Mean shortest-path length | 2.667 | 2.676 | +0.3% |
| Network diameter | 7 | 8 | +1 (+14.3%) |

**Legend.** Network topology before and after computational removal of the bone marrow endothelial node and its incident interactions. Mean shortest-path length and diameter were calculated within the largest weakly connected component, as described in the submitted manuscript. Percentage changes were calculated from the displayed values. Original Gephi outputs are shown in Supplementary Screenshots S11 and S12.

**Table 6. Within-corpus physiological comparison**

| Within-corpus comparison | n | Percentage |
|---|---|---|
| Physiological relationships | 33 | — |
| Mappable to final ontology | 32 | 97.0% |
| Shared directed pairs | 14 | 43.8% of 32 mappable relationships |
| Physiological-only direct pairs | 18 | 56.3% of 32 mappable relationships |
| Shared diabetic edges | 14 | 17.7% of diabetic network |
| Diabetic-network interactions without a physiological match | 65 | 82.3% of 79 diabetic interactions |
| Unmatched therapeutic/intervention interactions | 5 | 6.3% of 79 diabetic interactions |
| Unmatched non-intervention interactions | 60 | 75.9% of 79 diabetic interactions |

**Legend.** Comparison of corpus-derived physiological relationships with the frozen diabetic network after harmonisation to the final canonical node ontology. Physiological-only denotes absence of the corresponding direct source→target edge from the diabetic reconstruction and does not imply biological disappearance of the relationship.

**Table 7. Biological categories of unmatched non-intervention interactions**

| Biological category | Unmatched interactions |
|---|---|
| Inflammatory myelopoiesis | 12 |
| Endothelial–vascular niche | 12 |
| Epigenetic/transcriptional memory | 9 |
| Neural–circadian mobilisation | 7 |
| HSPC maintenance/mobilopathy | 7 |
| Sphingolipid remodelling | 5 |
| Upstream stress | 3 |
| Adipogenic remodelling | 3 |
| Systemic modifier | 2 |
| **Total** | **60** |

**Legend.** Distribution of the 60 non-intervention interactions without a matched pair in the partial physiological reference. Categories are based on biological annotation and are distinct from graph-derived communities. Unmatched status does not establish specificity to diabetes.

**Table 8. Functional rewiring of shared physiological–diabetic directed relationships**

| Directed relationship | Physiological state | Diabetic state |
|---|---|---|
| Endothelium → CXCL12 | Produces/regulates CXCL12 | Downregulates CXCL12 |
| CXCL12 → CXCR4 | Activates physiological retention signalling | Decoupled/desensitised |
| SCF → HSPCs | Maintains HSPCs | Deficient maintenance |
| Endothelium → HSPCs | Maintains/regulates HSPC fate | Loses angiocrine support |
| Endothelium → Notch | Activates angiocrine support | Suppresses angiocrine cues |
| Notch → HSPCs | Supports maintenance and lineage control | Fails to maintain normal fate |
| Pericytes → endothelium | Supports vascular niche integrity | Detachment/dysfunction |

**Legend.** Examples of shared directed source→target pairs whose encoded biological meaning differed between the corpus-derived physiological reference and diabetic network. Structural preservation of an edge was therefore distinguished from preservation of its functional interpretation.

**Table 9. Topological sensitivity to addition of the endothelium→SCF relationship**

| Metric | Frozen network | Added edge Endothelium → SCF | After endothelial deletion | Augmented + endothelial deletion |
|---|---|---|---|---|
| Weak components | 18 | 18 | 19 | 19 |
| Largest component (nodes) | 45 | 45 | 43 | 43 |
| Mean shortest-path length | 2.779 | 2.754 | 3.792 | 3.792 |
| Diameter | 5 | 5 | 10 | 10 |
| Global efficiency | 0.218 | 0.220 | 0.170 | 0.170 |

**Legend.** Comparison of the frozen network and the endothelium→SCF-augmented sensitivity network before and after endothelial deletion. The augmented edge was introduced only into the secondary sensitivity graph; the frozen 62-node/79-edge network was unchanged. **The** augmented baseline contained 62 nodes and 80 edges. Both deletion graphs contained 61 nodes and 62 edges because the added endothelial edge was removed with its source node.

**Table 10. Graph-constrained AI-assisted interrogation-fidelity analysis**

| Q | Property tested | Graph-defined reference | Outcome |
|---|---|---|---|
| Q1 | Route recovery | SNS→β-adrenergic→CXCL12→CXCR4→HSPCs; 4 edges | Route recovered |
| Q2 | Route recovery | ROS→RhoA/ROCK→Akt/eNOS→endothelium→Notch→HSPCs; 5 edges | Route recovered |
| Q3 | Route recovery | S100A8/A9→RAGE→NF-κB→M-CSF/GM-CSF→CMP/GMP→myeloid bias; 5 edges | Route recovered |
| Q4 | Route recovery | Diabetes→adipocytes→MCP-1/CCR2→macrophages→systemic inflammation; 4 edges | Route recovered |
| Q5 | Unsupported shortcut | Direct endothelium→SCF absent; endothelium→ANGPT1→SCF→HSPCs encoded | Shortcut rejected; valid alternative recovered |
| Q6 | Semantic fidelity | Endothelium→CXCL12→CXCR4→HSPCs | Downregulation, decoupling or desensitisation and weakened retention preserved |
| Q7 | Directionality | Proposed CXCR4→CXCL12 | Rejected; encoded direction CXCL12→CXCR4 |
| Q8 | Direct vs indirect reachability | Direct endothelium→Akt/eNOS absent; endothelium→EGFR→Akt/eNOS encoded | Direct edge rejected; indirect route retained |
| Q9 | Directionality | Proposed RAGE→S100A8/A9 | Rejected; encoded direction S100A8/A9→RAGE |

**Legend.** Nine graph-constrained queries testing multistep route recovery, resistance to an unsupported shortcut, semantic preservation and edge-direction fidelity. Reference answers were defined by the frozen E001–E079 edge table; the analysis assessed fidelity to the encoded architecture rather than biological correctness of the graph.

## ACKNOWLEDGEMENTS

The author thanks the developers of contemporary large language models for providing analytical tools that enabled structured interrogation of a complex biomedical evidence corpus. The author also acknowledges the open-source computational biology community for the development and maintenance of the network analysis and Boolean modelling platforms cited in this work. The conception of the study, construction of the evidence corpus, analytical strategy, biological interpretation, hypothesis generation and all conclusions presented in this study are solely those of the author.

## AI TRANSPARENCY STATEMENT

Structured dialogue with LLMs supported evidence extraction, candidate synthesis and biological hypothesis generation under investigator supervision. The supplied publications, not model-generated assertions, constituted the evidential basis of reconstruction. Models generated candidate conclusions, which the investigator reviewed against the source literature before consensus integration. The author retains responsibility for final interpretation, manuscript content and all scientific conclusions.

## FUNDING

This research received no specific grant from any funding agency in the public, commercial or not-for-profit sectors.

## CONFLICT OF INTEREST

The author declares no competing financial interests.

## DATA AVAILABILITY

The node and edge tables, concordance summaries, prompts and post-reconstruction audits are provided in the supplementary files. Verbatim platform responses are available upon request.